\documentclass[aps,pre,preprint,superscriptaddress,showpacs,amsmath,amssymb,floatfix]{revtex4-1}

\usepackage{graphicx}  
\usepackage{dcolumn}   
\usepackage{bm}        
\usepackage{amsmath}   
\usepackage{amsfonts}
\usepackage{amssymb}
\usepackage{natbib}
\usepackage{xfrac}    
\usepackage{nicefrac} 
\usepackage{color}
\definecolor{lightblue}{rgb}{0.9,0.95,0.97}
\definecolor{darkblue}{rgb}{0,0,0.6}
\definecolor{darkred}{rgb}{0.6,0,0}
\definecolor{green}{rgb}{0.0,0.3,0}
\renewcommand{\vec}[1]{\mathbf{#1}}

\renewcommand{\equiv}{\Leftrightarrow}

\begin{document}

\title{Permanent shear localization in dense disordered materials due to microscopic inertia}
\author{Vishwas V Vasisht} 
\author{Magali Le Goff} 
\author{Kirsten Martens}
\author{Jean-Louis Barrat}

\affiliation{Univ.~Grenoble Alpes, CNRS, LIPhy, 38000 Grenoble, France}

\begin{abstract}
In this work we develop a theoretical framework for the localization of flow in the steadily flowing regime of sheared disordered solids with inertial dynamics on a microscopic scale. To this aim we perform rheology studies at fixed shear rate on a 3D model of dense disordered solid. Our particle based simulations reveal the existence of heterogeneous shear-profiles in the stationary flow under homogeneous driving conditions. To rationalize this result, we propose a continuum model that couples the dynamics of the local flow to the evolution of a kinetic temperature field. A linear stability analysis of this theory predicts the minimum system size necessary for the flow instability to develop. This prediction as well as the velocity profiles obtained from this continuum model are in good agreement with the results from the particle based simulations. 
\end{abstract}

\maketitle

\section*{Introduction}Disordered materials, ranging from pharmaceutical, cosmetic and food products to construction components, such as glass or cement, form an integral part of our everyday life. Instabilities formed during the flow of such materials are ubiquitous and have been widely studied in the last few decades \cite{schall2010shear, fielding2014shear, divoux2016shear}. These flow instabilities  manifest in terms of strong spatial inhomogeneities in the flow, e.g. in form of shear bands, even when the material is homogeneously driven. Although several different scenarios have been shown to lead to shear banding, the origin of the instability in dense disordered solids is still debated in the literature. In this work, we propose a quantitative theoretical framework for the formation of shear bands in materials where inertia plays a role in the microscopic dynamics. 

Among the different scenarios leading to permanent shear banding, the case of polymeric systems and wormlike micelles solutions is well understood. In these systems,
shear induced structuration leads to the coexistence of regions flowing at different shear rates, even when the system is driven homogeneously \cite{lerouge2009shear, olmsted2008perspectives, divoux2016shear,dhont2008gradient}. In the framework of continuum mechanics, these permanent shear bands have been understood as a consequence of a material instability, e.g.~resulting from a non-monotonic constitutive flow curve (macroscopic stress {\it vs.} applied shear rate). Shear banding was also reported in the case of dense disordered solids, that exhibit a yield stress \cite{bonn2017yield}, such as concentrated emulsions, foams and other dense colloidal suspensions \cite{schall2010shear, fielding2014shear, divoux2010transient}. Although there is lack of consensus on the origin of shear banding in such materials, in the case of shear history-dependent materials \cite{coussot2002coexistence, ragouilliaux2007transition, moller2008shear} and dense hard sphere suspensions \cite{besseling2010shear}, theoretical approaches based on coupling flow fields with either the micro-structure \cite{fielding2009shear, mansard2011kinetic} or concentration field \cite{besseling2010shear, gross2018shear} have been successful in predicting permanent shear bands.

The case of soft dense suspensions, where neither significant structural nor volume fraction inhomogeneities are observed \cite{ovarlez2013existence, bonn2017yield}, remains however unclear. The role of attractive or adhesive interactions, proposed to lead to shear banding, is debated in the literature \cite{becu2006yielding, ovarlez2013existence, chaudhuri2012inhomogeneous}. Alternatively, an intrinsic timescale for "restructuration" at the microscopic scale is suggested as a mechanism to induce local weakening leading to shear banding \cite{coussot2010physical, martens2012spontaneous}. However this mechanism has not yet been evidenced at the microscopic level. Recent works, looking at the role of inertia on the flow behavior of yield stress materials \cite{salerno2012avalanches, karimi2016role, karimi2017inertia,nicolas2016effects} have demonstrated rate-weakening mechanisms resulting in non-monotonic macroscopic flow curves. This effect has been rationalized by Nicolas {\it et al.} \cite{nicolas2016effects} in terms of kinetic heating of the system due to inertia, an approach in agreement with the idea of a granular temperature in the deformation of granular materials \cite{losert2000particle}. But although it has been shown that inertia on the microscopic scale indeed leads to non-monotonic flow curves, no evidence of shear localization has been found in these former studies.

In this work we show that kinetic heating due to inertia can indeed lead to local softening and hence shear banding. However, the coupling of the mechanical properties to an additional destabilizing field appears only as a necessary condition, as a minimum system size is required to observe a permanent flow instability. We propose a continuum model based on a kinetic temperature description that allows for the prediction of the development of shear instabilities for system sizes exceeding a critical size. We also find that the steady state profiles at different applied shear rates do not form a banded structure following a lever rule, a phenomenon that we can again rationalize using the same continuum description.

\section{Numerical evidence for a steady state instability}
The model disordered solid considered in this work is an assembly of particles at volume fraction $\phi=70\%$, interacting via a truncated and shifted Lennard-Jones potential \cite{weeks1971role} defined as $U(r) = 4\epsilon \left [(a_{ij}/r_{ij})^{12} - (a_{ij}/r_{ij})^{6} \right ] + \epsilon$ , for $r_{ij} \le 2^{1/6} a_{ij}$, else $U(r_{ij})=0$. Here $a_{ij} = (a_i+a_j)/2$ defines the distance between the center of particles $i$ (with diameter $a_i$) and $j$ (with diameter $a_j$) at contact and the unit energy $\epsilon=1$ for all particles. The diameters of the particles are drawn from a Gaussian distribution with a variance of 10\%. The initial sample system is prepared from a high temperature dense liquid cooled down to a low temperature, using a NVT Molecular Dynamics protocol. The samples are subjected to a shear deformation at a fixed shear rate $\dot{\gamma}$ using Lees-Edwards boundary conditions (LEBC) and solving the following equation of motion based on dissipative particle dynamics (DPD):
\begin{equation}
m \frac{d^2\vec{r}_i}{dt^2} = - \zeta \sum_{j(\ne i)} \omega(r_{ij}) (\hat{r}_{ij}\cdot \vec{v}_{ij})\hat{r}_{ij} - \triangledown_{\vec{r}_i} U
\end{equation}
\noindent where the first term in the right hand side (RHS) is the damping force which depends on the damping coefficient $\zeta$ and $m$ is the mass of the particle. The relative velocity $\vec{v}_{ij} = \vec{v}_j - \vec{v}_i$ is computed over a cut-off distance $r_{ij} \le 2.5 a_{ij}$, with the weight factor $\omega(r_{ij})=1$. The second term in the RHS is the force due to interactions between particles. As a measure of the extent of over damping, we define an inertial quality factor $Q=\tau_\mathrm{damp}/\tau_\mathrm{vib}$, where $\tau_\mathrm{damp}=m/\zeta$ is the viscous timescale and $\tau_\mathrm{vib}=\sqrt{m a^2 / \epsilon}$ defines an elastic timescale. The inertial quality factor measures the number of inertial oscillation within the damping time. The shear rate in our simulations is defined in the units $\tau_\mathrm{vib}^{-1}$. Similar to 2D simulation results in \cite{nicolas2016effects}, we find in our simulations that an inertial quality factor $Q=1$ leads to a monotonic flow curve well fitted by an Herschel-Bulkley (HB) law $\sigma(\dot{\gamma}) = \sigma_y + A \dot{\gamma}^n$, with $\sigma_y \approx 2.5$ the yield stress, a coefficient $A \approx 16.5$ and the HB exponent $n\approx0.5$.

\noindent{\bf Non-monotonic flow curve: } We investigate the inertial dynamics of the jammed system under shear by considering the inertial quality factor $Q=10^4$. We perform finite shear rate simulations for a shear rate range of $10^{-4} \tau_\mathrm{vib}^{-1} \le \dot{\gamma} \le 0.5 \tau_\mathrm{vib}^{-1}$ and compute the flow curves for various sizes and geometries (maintaining a constant volume fraction $\phi=70\%$) (see Fig.~\ref{fig1}). In all our simulations [x,y,z] dimensions refers to flow, gradient and vorticity directions respectively. For the cubic system ($L_x=L_y=L_z=42 a$), with around $100K$ particles, we observe a non-monotonic flow curve which is attributed to the under-damped dynamics \cite{nicolas2016effects}. The flow curve has a minimum at $\dot{\gamma}\approx 0.1 \tau_\mathrm{vib}^{-1}$. Hence one could expect a flow instability in the region of shear rates $\dot{\gamma}<0.1\tau_\mathrm{vib}^{-1}$, where the flow curve exhibits a negative slope. Similar to the observations of Nicolas {\it et al.} \cite{nicolas2016effects} we did not find any shear instabilities in the velocity profiles. These results remain unchanged even for a non-cubic geometry, where $L_x=L_z=42 a; L_y=120 a$. With further increase in $L_y$ we observe a lowering of the steady state stress for shear rates below $\dot{\gamma}= 0.1\tau_\mathrm{vib}^{-1}$, while the flow curve remains unchanged in the positive sloped region, suggesting the appearance of a new flow regime for large systems in the regime $\dot{\gamma} < 0.1\tau_\mathrm{vib}^{-1}$.

\noindent {\bf Permanent shear instabilities: }We show, in Fig.~\ref{fig1}(b)-(c), for $L_y=360 a; \dot{\gamma}= 10^{-2} \tau_\mathrm{vib}^{-1}$, that the system exhibits permanent shear heterogeneities along the gradient direction in the flow regime mentioned above. The local shear rates and stresses were computed after the system had been sheared for around $\gamma=60$ (or 6000\% strain) and averaged over strain window of $\Delta \gamma=0.20$. It is clear that the system has reached a steady state both from the load curve (see SI) and from the homogeneous stress profile displayed in Fig.~\ref{fig1}(c).
The local shear rate profile in Fig.~\ref{fig1}(b) is clearly showing the feature of steady state shear localization. To our knowledge this is the first time such a formation of shear localization is reported for a dense disordered system without walls (periodic boundary conditions).
These set of simulation results indicate that the instability resulting from a negatively slope in the flow curve can only be observed above a minimum system size that can accommodate the corresponding instability.

In the following, we propose a continuum model based on the ideas of ref. \cite{nicolas2016effects}, in order to predict the minimum length scale required to allow for such an inertia-induced flow instability.

\section{Mechanism for inertia-induced instability}

\noindent{\bf Kinetic temperature due to inertia: } If we consider that inertia is introducing a kinetic temperature $\tilde{T}$ to the athermal dynamics, quantified through an excess of kinetic energy \cite{losert2000particle,nicolas2016effects, karimi2016role}, one can interpret the non-monotonic flow rheology of athermally driven inertial systems $\sigma(\dot{\gamma};Q,T=0)$ effectively as an over-damped rheology at a finite shear-rate dependent temperature $\sigma(\dot{\gamma}; Q=1, \tilde{T}(\dot{\gamma}))$ (see Fig.~\ref{fig2}(a)). We know that in over-damped situations the athermal flow rheology of simple yield stress fluids is usually well described by a Herschel-Bulkley type monotonous relationship between the shear stress and the applied shear rate $\sigma_0(\dot{\gamma})=\sigma(\dot{\gamma};T=0)=\sigma_y + A\dot{\gamma}^n$ with $n\approx 0.5$. The temperature dependence of the pre-factor in this rheological relation can be neglected, $A(\tilde{T})\approx A(T=0)$ (see Fig.~\ref{fig2}(b). The functional form of the first correcting term in the macroscopic flow curve for small but finite temperatures has been worked out in previous works but is valid only at large enough shear rates \cite{johnson2005universal, chattoraj2010universal}. Lacking a specific form of the temperature dependence of the flow curve, that can be sensitive to the details of the modeling and the interaction potentials, we only assume a formal dependence of the local stress on the local kinetic temperature.

\noindent {\bf Continuum model: } To fix the notations we assume in the following a specific driving protocol at a fixed shear rate $\dot{\gamma}=\partial v_x/\partial y$ in a 3d planar geometry (flow direction x, gradient direction y, and vorticity direction z). Using the continuity equations for momentum and energy, we derive the time evolution equations for the shear component of the velocity $v_x(y,t)$ and the kinetic temperature $\tilde{T}(y,t)$ as \cite{katsaounis2017emergence}:

\begin{eqnarray}
 \rho \frac{\partial v_x}{\partial t} &=& \frac{\partial \sigma}{\partial y} = \frac{\partial \sigma}{\partial \tilde{T}}\frac{\partial \tilde{T}}{\partial y} + \frac{\partial \sigma}{\partial \dot{\gamma}}\frac{\partial^2 v_x}{\partial y^2} \label{eq:continuum_model_1}\\
  c_V \frac{\partial \tilde{T}}{\partial t} &=& \lambda_T \frac{\partial^2 \tilde{T}}{\partial y^2}  + \frac{\partial v_x}{\partial y}\sigma\left(\frac{\partial v_x}{\partial y},\tilde{T}\right) - \frac{c_V}{\tau} \tilde{T}\;.\label{eq:continuum_model_2}
\end{eqnarray}
\noindent where $\rho$ is the system density, $c_V$ the volumetric heat capacity, $\lambda_T$ the thermal conductivity and $\tau$ the typical time to remove the kinetic energy (originating from the external shear) by the thermostat (at zero temperature). As a test of our continuum model we first obtain from the above equations the relation between temperature and shear rate in the stationary state
($\partial \sigma/\partial t = 0$; $\partial \sigma/\partial y = 0$; $\partial \tilde{T}/\partial t=0$; $\partial \tilde{T}/\partial y=0$) and uniform flow limit ($\partial \dot{\gamma}/\partial y=0$), given by
\begin{equation}\label{eq:T_steady_state}
	\tilde{T} = \frac{\tau}{c_V} \sigma(\dot{\gamma}) \dot{\gamma}
\end{equation}
\noindent In simulations one can measure $\lambda_T$ and $c_V$, which are system properties fixed by the interaction potential and $\tau$ is fixed by the dissipation coefficient $\zeta$. For $Q=10^4$ and $L_x=L_y=L_z=42 a$ we measure, as a function of $\dot{\gamma}$, the kinetic temperature $\tilde{T}$ as $0.5 <m V_z>^2$, where $V_z$ is the velocity of a particle in the neutral vorticity direction, which is not influenced by the affine flow. In Fig.~\ref{fig2}(c), we show $\tilde{T}$ measured from simulations as well as the one obtained from Eq.~(\ref{eq:T_steady_state}). and we find a good match between the two, showing that the kinetic temperature emerging from the continuum model is well describing the simulations in the homogeneous flow regime.

\noindent {\bf Stability analysis and system size dependence: } Next we perform a linear stability analysis for small perturbations of the homogeneous flow solution $v_x(y,t)=\dot{\gamma}y+\delta v_x(y,t)$ and of the constant kinetic temperature field $T(y,t)=T_0+\delta T(y,t)$ using the ansatz $\delta v_x(y,t) = \tilde{v}_x (t) \exp\{-iky\}$ and $\delta T(y,t) = \tilde{T} (t)\exp\{-iky\}$. Linearizing the equations and solving for the characteristic polynomial, the product of the eigenvalues $\lambda_1$ and $\lambda_2$ yields
\begin{eqnarray}\label{eq:lin-sta}
 \lambda_1\lambda_2=\frac{\lambda_T}{\rho c}\frac{\partial \sigma}{\partial \dot{\gamma}} k^4 + \frac{1}{\rho c}\left(\sigma\frac{\partial \sigma}{\partial T}+\frac{c}{\tau}\frac{\partial \sigma}{\partial \dot{\gamma}}\right) k^2\;.
\end{eqnarray}
For the homogeneous flow to be stable we need both eigenvalues to be negative and thus the above expression to be positive. The first term on the right hand side is indeed positive and thus stabilizing, since we started from a monotonic flow curve assumption in the over-damped limit. The sign of the second term will however depend on the competition of two contributions, one stabilizing related to the efficiency of the thermostat and one destabilizing related to the sensitivity of the flow curve to the increase in kinetic temperature. When this last term becomes predominant in the case of weakly damped systems, we expect the homogeneous flow solution to become unstable for large wavelength perturbations. In that case the critical wavelength $\ell_c=2\pi/k_c$ can be estimated as

\begin{eqnarray}\label{eq:cri-len}
 \ell_c&=&2\pi\sqrt{\lambda_T}\left(-\sigma\frac{\partial_T \sigma}{\partial_{\dot{\gamma}} \sigma}-\frac{c}{\tau}\right)^{-\frac{1}{2}}\;,
\end{eqnarray}
where we used a simplified notation for the partial derivative operator $\partial_\bullet f = \partial f/\partial \bullet$. This discussion points to the existence of a critical system size $\ell_c$ above which a shear instability can develop.

We compute $\ell_c$ using Eq.~(\ref{eq:cri-len}) inputting the values of numerically measured parameters as well as estimates of $\partial_T \sigma$ and $\partial_T{\dot{\gamma}}$ from the simulations (details in SI). In the parametric plot of Fig.~\ref{fig3}, we show $L_y$ as a function of $\dot{\gamma}$. $\ell_c$ computed from eq. \ref{eq:cri-len} separates two different flow regimes.
The small system regime (red) corresponds to the homogeneous flow being stable, which becomes unstable for large systems (blue). The symbols represent numerical simulations data where we find either an homogeneous flow (red triangles) or a localized flow (blue circles). We find a good match between the prediction from the continuum model and numerical simulations. These results indicate that our model can predict quantitatively the onset of the flow instability observed in microscopic simulations. The properties of the heterogeneous flow regime are discussed in the next section.

\section{Steady state profiles and lack of lever rule}
We now discuss the dependence of the flow profiles on the applied shear rate. In usual shear banding scenarios \cite{divoux2016shear}, if the average stress is homogeneous in the system, one generally expects the maximum shear rate achieved in the system to be independent of the applied shear rate, as well as an increasing band width with an increase in shear rate, so as to conserve the applied shear rate (so called 'lever rule'). In Fig.~\ref{fig4}(a), we show the local shear rate profiles obtained from molecular simulations for different applied shear rates for a given system size (with $L_y>\ell_c(\dot{\gamma})$). The profiles show that, even though the width of the band increases with an increase in shear rate, there is a clear absence of one chosen maximum shear rate. Instead, the profiles exhibit a continuous interface, without a clear plateau associated with the flowing region. Similar features are observed in the flow profiles computed from the model (Fig.~\ref{fig4}(b)). To obtain the flow profiles of Fig.~\ref{fig4}(b) we integrate the continuum model in steady state (Eq.~(\ref{eq:continuum_model_2}) with LHS term equal to 0) using the shooting method for single-banded profiles. To this aim, we assume a simple constitutive relationship relating the shear stress to the shear rate and kinetic temperature, based on a Herschel-Bulkley description at a given temperature:
\begin{eqnarray}
\sigma = \sigma_y + A\dot{\gamma}^n - B\tilde{T}^\alpha
\label{eq:flowcurve}
\end{eqnarray}
\noindent where $A$ and $n$ are fitting parameters for the molecular simulation data at zero temperature, $B=2.3$ and $\alpha=0.3$ describe the decrease in stress with kinetic temperature. Using Eq.~(\ref{eq:T_steady_state}), relating $\tilde{T}$ to $\dot{\gamma}$ for an homogeneous flow, one can solve the implicit equation given by Eq.~(\ref{eq:flowcurve}), which leads to a non monotonic constitutive flow curve (see SI). This expression is not the best fit to the MD data but constitutes the simplest form to qualitatively reproduce the main features of the flow curve, with a minimum located at ($\dot{\gamma}_\mathrm{min}$, $\sigma_\mathrm{min}$) close to the values of MD simulation.

\noindent{\bf Rationalizing the flow profiles: } A stationary localized flow corresponds to a situation of coexistence between flowing and nearly immobile regions, with a homogeneous stress profile. The determination of the flow profile is analogous to the determination of interfacial profiles in phase coexistence problems, and can be described using  a classical mechanical analogy \cite{rowlinson1982molecular,oxtoby1982jcp}. The steady state temperature $T(y)$ obeys equation 3 with $\partial_t \tilde{T}=0$, and this equation can be interpreted as describing the trajectory $\tilde{T}(y)$ of a fictitious particle in an external potential, in one dimension: 

\begin{eqnarray}
\lambda_T \frac{d^2 \tilde{T}}{d y^2}  = -\frac{d U(\tilde{T})}{d \tilde{T}}.
\end{eqnarray}
\noindent In this mechanical analogy, $\tilde{T}$ corresponds to the position of the particle and $y$ to time. 

This effective potential is computed by integrating Eq.~(\ref{eq:continuum_model_2}) in steady state with respect to $\tilde{T}$ at a fixed value of the stress $\sigma$: 
$U(T) = \sigma \int_0^T \dot{\gamma}(\sigma,T') dT' -c_VT^2/2\tau$, where the function $\dot{\gamma}$ is  expressed using the flow curve given in Eq.~(\ref{eq:flowcurve}) as 
\begin{equation}
\dot{\gamma}(\sigma,\tilde{T}) = A^{-1/n}(\sigma-\sigma_y+B\tilde{T}^\alpha)^{1/n}\  \mathrm{if} \ \sigma > \sigma_y -B\tilde{T}^\alpha; 
\end{equation}
\begin{equation}
\dot{\gamma}(\sigma,\tilde{T}) = 0\  \mathrm{otherwise} \ 
\end{equation}

The resulting potential is displayed in Fig.~\ref{fig5}(a) for various values of the stress, and will define four different regimes for the possible trajectories. We first note that, by construction, extrema of the effective potential correspond to temperatures (and shear rates) that are solution of the set of equations \ref{eq:T_steady_state} and \ref{eq:flowcurve}, which describe homogeneous flow. Below a minimum stress $\sigma_\mathrm{min}$, which will correspond to the minimum of the actual flow curve, the only extremum is obtained for  $\tilde{T}=0$, which implies $\dot{\gamma}=0$. This situation is depicted by the brown solid line in Fig.~\ref{fig5}(a). For $\sigma < \sigma_\mathrm{min}$, no flow is possible. The second simple regime is the high stress one, $\sigma > \sigma_y$. In this case (not shown in the figure) the extremum at $T=0$ disappears,  $U(\tilde{T})$ has a finite, positive slope at $\tilde{T}=0$ and  a single maximum at high temperature. This maximum corresponds to homogeneous flow  in the high shear rate, high stress regime. A richer situation is observed for intermediate stresses, $\sigma_\mathrm{min}<\sigma<\sigma_{y}$. Here $U(\tilde{T})$ has three extrema: a maximum at $\tilde{T}=0$, a second maximum at  $\tilde{T}_\mathrm{max}$, and an intermediate minimum at $0 <\tilde{T}_\mathrm{min}<\tilde{T}_\mathrm{max}$. Possible interfacial profiles correspond to oscillations of $\tilde{T}$ around the minimum, with the "period" of the oscillation being equal to the size of the system. Here, two cases must be distinguished, and we focus first on the one that corresponds to the profiles effectively observed in simulations, with $U(\tilde{T}_\mathrm{max}) > U(0)$, as illustrated by the blue curve in figure \ref{fig5}(a).  In this case, an oscillation can be obtained for values of the potential energy $U$ between $U(T_\mathrm{min})$ and $U(0)$. The period of the oscillation will have a value that starts from a minimum, nonzero value for the smallest energies in the vicinity of $U(T_\mathrm{min})$, where the oscillation is harmonic. This minimum period defines the critical system size, below which no interface can be observed and the system remains in an homogeneous state. We have checked that the corresponding value of $\ell_c$ obtained from this analysis coincides with the one obtained from the linear stability analysis in Eq.~(\ref{eq:cri-len}). As the value of the energy increases towards $U(0)$, the period of oscillation increases and becomes infinite at $U(0)$, where the trajectory spends a short time at a finite temperature and most of the time near $\tilde{T}=0$. This corresponds to a narrow sheared layer coexisting within a broad non-flowing part, as illustrated by the horizontal dashed blue line in the figure, where the point marks the maximum temperature $\tilde{T}_\mathrm{flow}$ inside the destabilized region.  The locus of $\tilde{T}_\mathrm{flow}$ as the stress varies is indicated by the blue line in \ref{fig5}(b), where the shaded blue region corresponds to the values of temperatures that can exist within the profile. Clearly the value of $\tilde{T}_\mathrm{flow}$, which can be defined by $U(\tilde{T}_\mathrm{flow})= U(0)$, depends on the applied stress. As a result, no lever rule is expected: the total shear rate increases as stress increases not only by broadening the sheared region, but also by increasing temperature and strain rate inside the flowing region.

The above described regime, which accounts well for the observations made in the atomistic simulations, is  observed for values of the stress $\sigma^* <\sigma <\sigma_y$. At $\sigma^*$, $U(\tilde{T}_\mathrm{max}) =U(0)$, and for $\sigma_\mathrm{min}< \sigma < \sigma^*$, $U(\tilde{T}_\mathrm{max}) < U(0)$. As a result, oscillating trajectories will exist between a small temperature and $T_\mathrm{max}$, with most of the "time" being spent in the vicinity of $T_\mathrm{max}$. This situation is illustrated with the purple lines in figure \ref{fig5}(a), and would correspond to a broad region at high shear rates coexisting with a narrower immobile layer. 
We have not observed this situation in the molecular dynamics simulations:  this is not surprising, as it  corresponds to a very narrow range of stresses,  approaching the minimum of the flow curve, where the critical size for the flow instability to develop becomes increasingly large (see Fig.~\ref{fig3}). 

\section{Conclusion}
Mathematical descriptions for the phenomenon of shear localization in the steady state flow of non-newtonian fluids usually invoke linear instabilities induced by spatial fluctuations of a field coupled to the flow (eg. local concentration \cite{dhont2008gradient,olmsted2008perspectives}, local temperature \cite{dafermos1983adiabatic, katsaounis2017emergence} or local microstructure \cite{dhont2008gradient}). Non-linear terms in these continuum descriptions ensure the existence of bounded stationary profiles in the long time limit. Although this picture appears very general, there is still a lack of quantitative models especially in the framework of yield-stress materials, that could be directly compared to experiments or simulations. 

\begin{figure*}[t]
\centering
\includegraphics[width=1.0\columnwidth, clip]{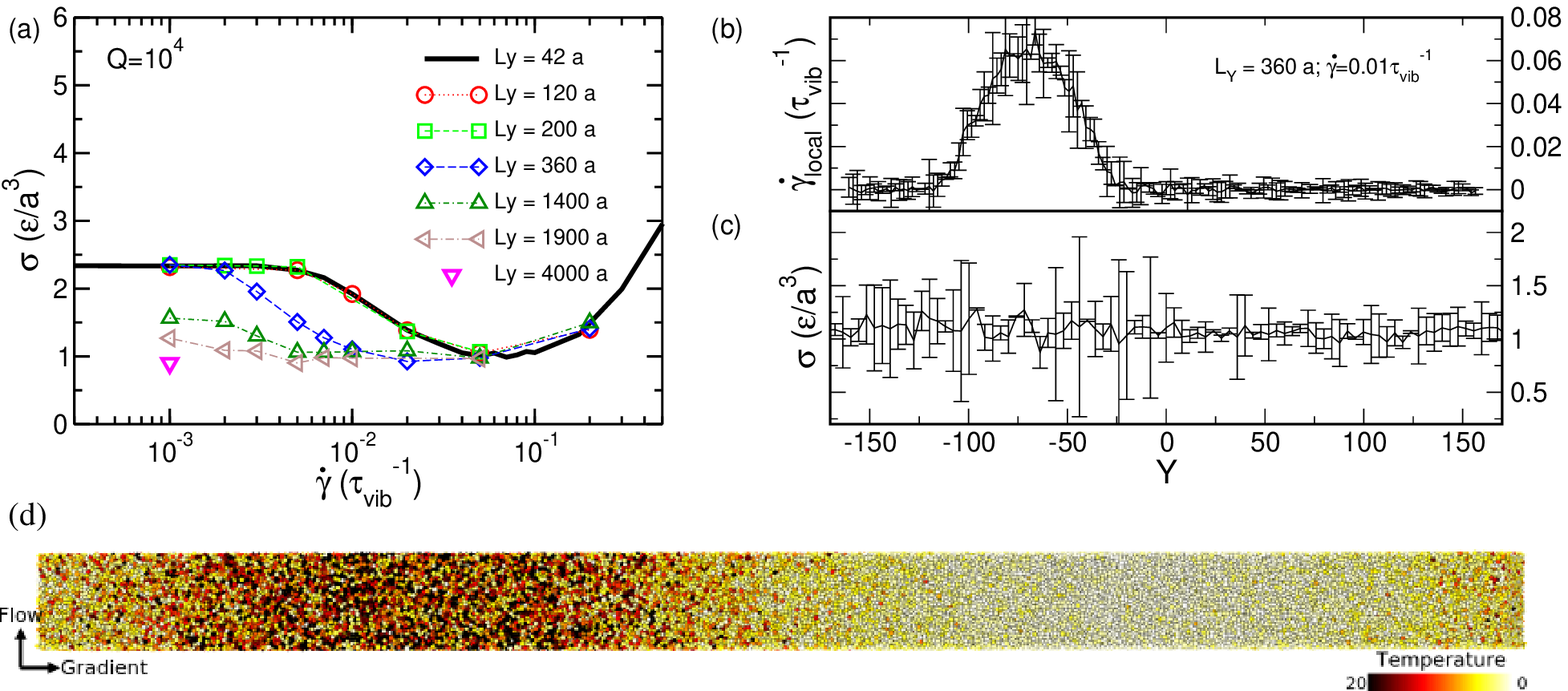}
\caption{(a) Flow curve showing steady state stress as a function of applied shear rate of athermal underdamped system ($Q=10^4$) for different system sizes. (b) Steady state local shear rate profile as a function of gradient direction, averaged over a strain window of $\Delta{\gamma}=0.2$, computed for a system size $N=240731$ and $L_Y=360 a$ at $\dot{\gamma}=10^{-2}\tau_\mathrm{vib}^{-1}$. (c) Corresponding stress profile. (d) 3D rendering of a configuration from simulation trajectory depicting shear localization ($Q=10^4$, $N=240731$, $L_Y=360 a$, $\dot{\gamma}=10^{-2}\tau_\mathrm{vib}^{-1}$). Hot coloring scheme, based on the kinetic temperature $\tilde{T}$, varies between white ($\tilde{T}=0 \epsilon/k_B$) to black ($V_x=20 \epsilon/k_B$).}
\label{fig1}
\end{figure*}

\begin{figure}
\centering
\includegraphics[width=1.0\columnwidth, clip]{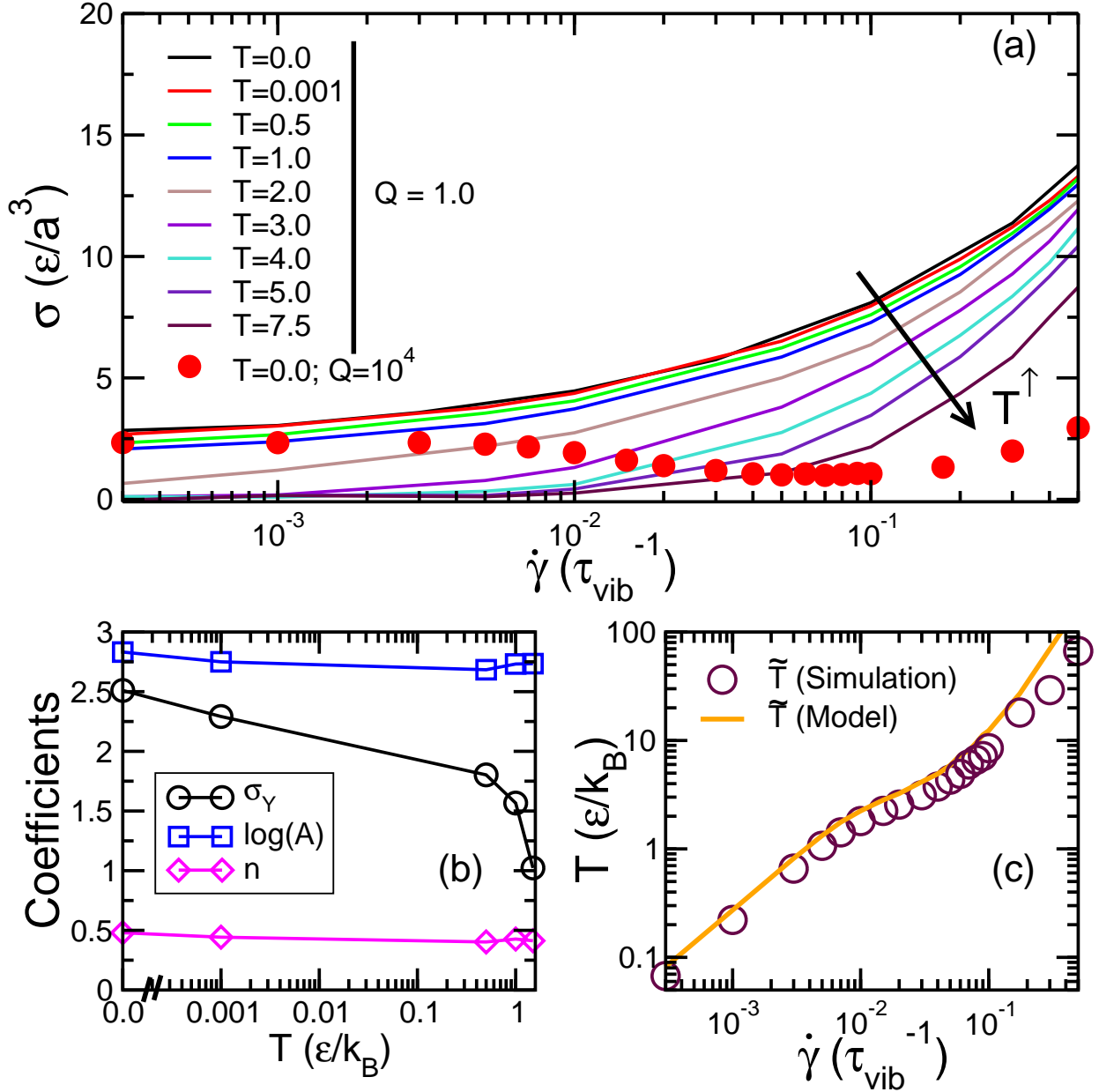}
\caption{(a)Flow curves obtained from over-damped simulations ($Q=1$) performed at finite temperature (solid lines) for a system size of $N=10^4$ particles. The filled circles represent the flow curve from under-damped simulation ($Q=10^4$, $N=100K$) at T = 0. (b) Temperature dependence of yield stress ($\sigma_Y$), coefficient (A) and exponent (n) obtained from fitting the over-damped flow curves with the Herschel-Bulkley equation $\sigma(\dot{\gamma}) = \sigma_y + A \dot{\gamma}^n$. (c) Comparison of kinetic temperature $\tilde{T}$ measured in simulations of under-damped ($Q=10^4$) athermal simulations with $\tilde{T}$ predicted from the continuum model (Eq.~(\ref{eq:T_steady_state})), for different shear rates.}
\label{fig2}
\end{figure}

The strength of this work is to single out a specific destabilizing field, and to propose a truly quantitative description by inferring all parameters of our proposed continuum model from particle based simulations. The case we study here is the one of shear-weakening induced by inertial dynamics at the microscopic scale. We show that this minor change in the dynamics can indeed lead to the phenomenon of shear-localization. The corresponding destabilizing field in the continuum model is defined through a local kinetic temperature. Contrary to effective fields, or parameters that enter many other continuum descriptions \cite{shi2007evaluation,manning2007strain,fielding2014shear,hinkle2017coarse}, the kinetic temperature has the advantage of having a clear microscopic definition, and is thus easily measurable in particle based simulations (see Fig.~\ref{fig1}(d)) and experiments \cite{losert2000particle}. 

Using this description it becomes possible to quantitatively predict the appearance of shear localization for systems larger than a critical size $\ell_c$. Beyond the linear instability analysis, we show that the qualitative features in the stationary profiles match well between particle simulations and the continuum model. Notably, the mathematical description allows to understand why the stationary profiles do not exhibit a simple band but a more complex continuous profile, leading to a lack of lever rule.

Our model has the potential to predict the overall dynamics including the stationary profiles if we had a better understanding together with an analytic expression for the effect of temperature on rheological flow curves. This would also allow for the study of transient features such as coarsening dynamics, both in microscopic simulations and from a theoretical standpoint, to understand the mechanisms for selection of specific non-linear solutions in the model. But it seems that the exact functional form of the flow curve may depend on simulation details and it might not be an easy task to derive such a generic expression \cite{nicolas2016effects}.

The aim of this work was to study in detail how a minimal ingredient can lead to permanent shear bands in soft amorphous solids (shear weakening induced by inertial dynamics). However, one has to note that in many of the systems where inertia is playing a role, there are also other ingredients at play known to produce flow instabilities, like for example friction in granular materials \cite{dijksman2011jamming,degiuli2017friction}. The emergence of hysteresis and shear bands in these systems could thus result from a complex interplay between the different mechanisms involved in the dynamics and potential candidates for flow instabilities. Following a similar approach as the one suggested in the present work, one expects that a complete continuum description should couple the stress dynamics to several destabilizing fields, of which the one of the local kinetic temperature might play an important role, as demonstrated in this work.

{\it Acknowledgements --}
J.-L.~B.~and V.~V.~V.~acknowledge financial support from ERC grant  ADG20110209  (GLASSDEF). 
K.~M.~acknowledges financial support from ANR Grant No. ANR-14-CE32-0005 (FAPRES) and CEFIPRA Grant No. 5604-1 (AMORPHOUS-MULTISCALE). J.-L.B.
is supported by IUF. Further we would like to thank Peter Olmsted, Guillaume Ovarlez and Romain Mari for fruitful discussions. 

\bibliography{biblio_shear_bands}
\clearpage
\newpage

\begin{figure}[t]
\centering
\includegraphics[width=1.0\columnwidth, clip]{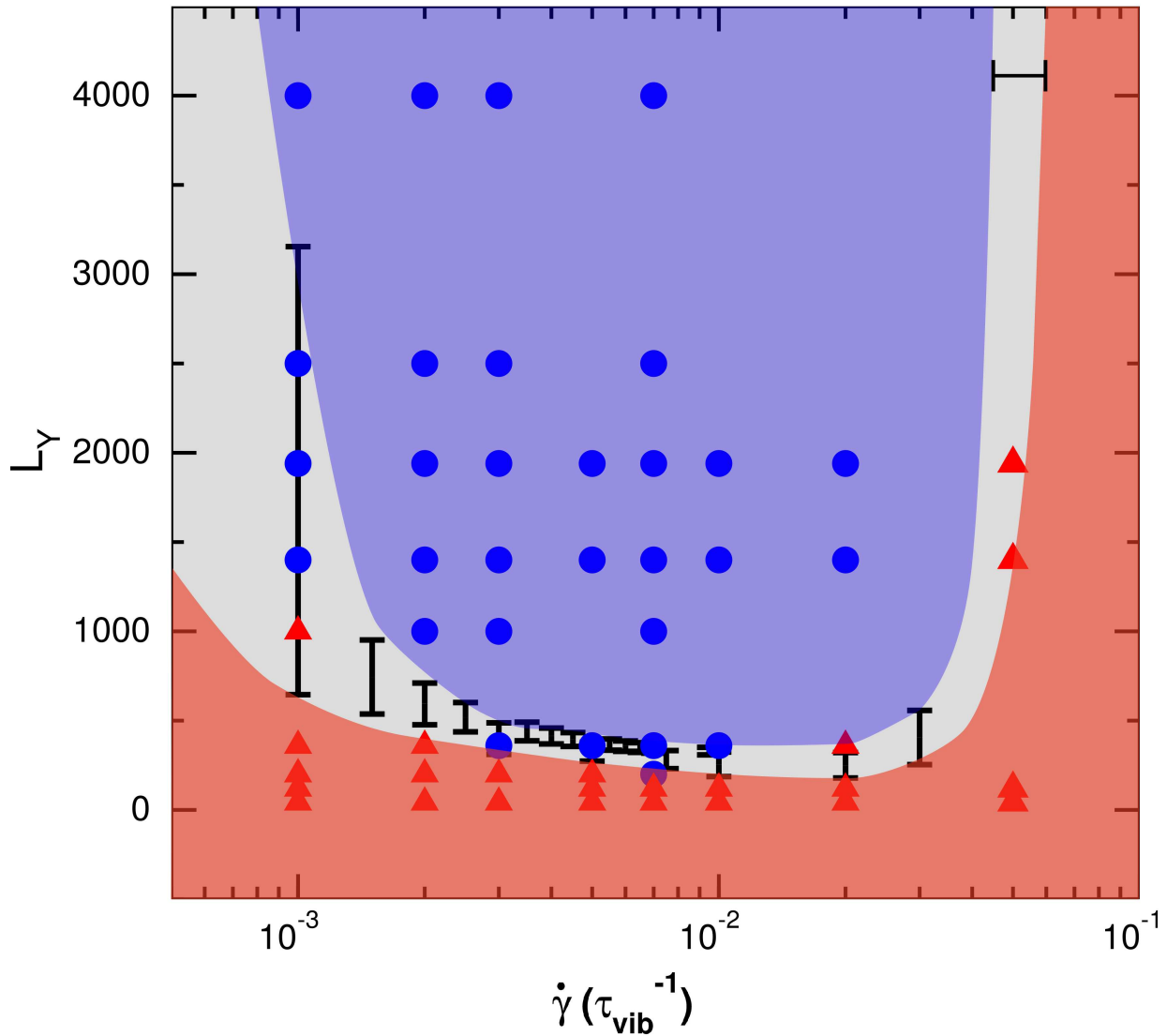}
\caption{System size (represented by the length of gradient dimension) as a function of shear rate. The symbols depicts the state points at which simulations were performed. Red triangle represent the state points where we find homogeneous flow and blue circles represent the state points where we find steady state shear instability. The value of $\ell_c$  computed from Eq.~(\ref{eq:cri-len}) is used to define the area coloring, with the red region representing the stable homogeneous flow and the blue region for observing flow instability. The method used to estimate $\ell_c$ (and error bars) is discussed in SI. The horizontal error bar corresponds to the error on the estimate of the minimum of the flow curve.}
\label{fig3}
\end{figure}

\begin{figure}[t]
\centering
\includegraphics[width=1.0\columnwidth, clip]{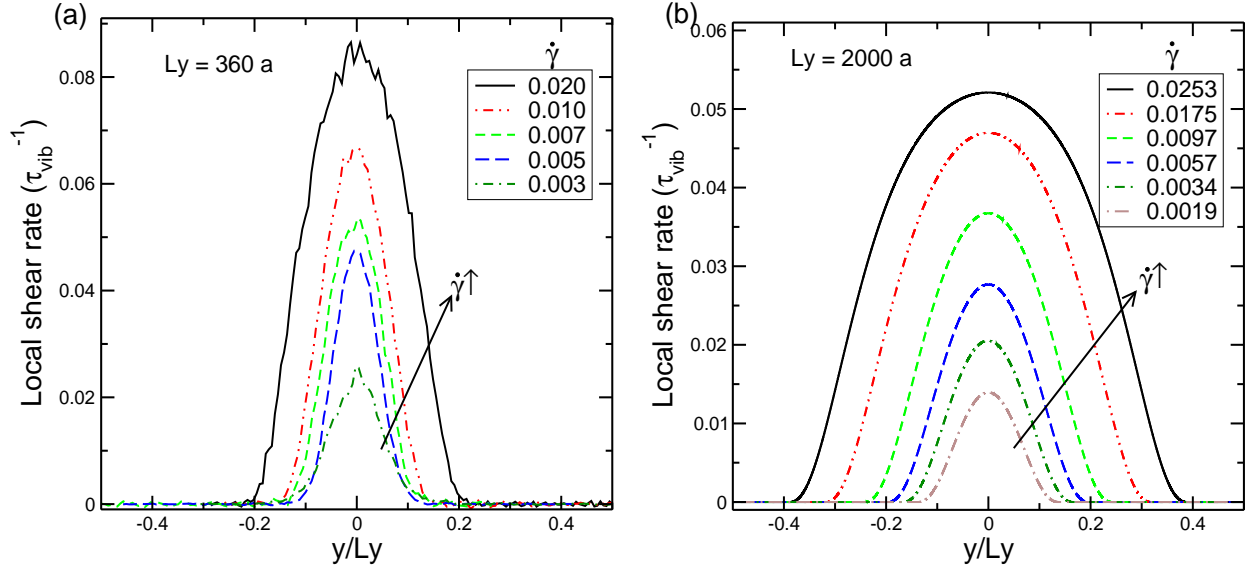}
\caption{Local shear rate profiles obtained from (a) molecular simulations for different applied shear rates (for a system dimension $L_x=L_y=22 a$ and $L_y=360 a$ and (b) continuum model (for a system size $L_y = 2000 a$), for different stresses using the shooting method .}
\label{fig4}
\end{figure}

\begin{figure}
\centering
\includegraphics[width=1.0\columnwidth, clip]{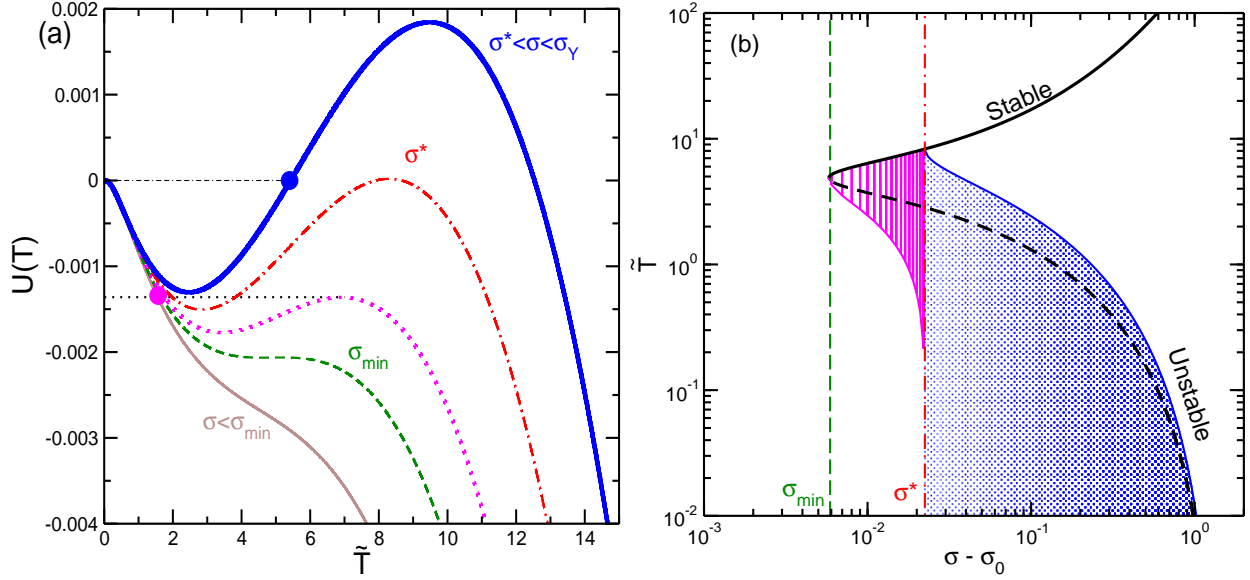}
\caption{(a) Effective potential $U(\tilde{T})$ as a function of $\tilde{T}$ for various values of stress $\sigma$. Brown solid line: no flow solution for $\sigma<\sigma_\mathrm{min}$. Dashed green line: Onset of a second maximum of $U(\tilde{T})$ for $\sigma=\sigma_\mathrm{min}$. Magenta dotted line: typical $U(\tilde{T})$ curve in the regime $\sigma_\mathrm{min}<\sigma<\sigma^*$ exhibiting three extrema, and possible shear localization with a wide flowing region (see text). Red dashed and dotted curve: $\sigma=\sigma^*$, stress at which the two maxima of $U(\tilde{T})$ have equal height. Blue solid line: typical $U(\tilde{T})$ curve in the regime $\sigma^*<\sigma<\sigma_y$ exhibiting three extrema, and possible shear localized flow with a wide arrested region (see text) and a flowing region with a continuous profile (see text). (b) Temperature-stress diagram. (x-axis shown as $\sigma-\sigma_0$, $\sigma_0$ being a value close to $\sigma_\mathrm{min}$ for a better display of the narrow region near the minimum of the flow curve). The solid blue line indicates the maximum temperature of the profile $T_\mathrm{flow}$ in the regime $\sigma^*<\sigma<\sigma_y$ and the solid magenta line indicates the minimum temperature in the system in the regime $\sigma_\mathrm{min}<\sigma<\sigma^*$ The blue and magenta shadowed area represent the coexisting temperatures in the two regimes allowing for shear instabilities.
}
\label{fig5}
\end{figure}

\clearpage
\newpage
\setcounter{figure}{0}

\noindent \textbf{\huge Supplementary Information}\\
\noindent{\bf Sample preparation protocol: } An initial FCC crystal is prepared at a chosen volume fraction of $0.70$ and a chosen set of dimensions for the simulations box given by $L_x$, $L_y$ and $L_z$. In all the simulations we fix $L_x=L_z$. This sample is melted at $T=5.0 \epsilon / k_B$ and equilibrated at the same temperature in a NVT ensemble for around 50K molecular dynamics (MD) steps, with a timestep of $\Delta t = 0.001$. By measuring the crystalline ordering, using the orientational order parameter $Q_6$, we make sure the system is disordered. Further this equilibrated melt is subjected to a systematic temperature quench from $T=5.0 \epsilon / k_B$ to to $T=0.001 \epsilon/k_B$ at a cooling rate $\Gamma=5 \cdot 10^{-3} \epsilon/(k_B \tau_0)$. After the system reaches $T=0.001 \epsilon/k_B$, we perform an energy minimization using a conjugate gradient method to take the system to the zero temperature limit.\\

\begin{figure}[h]
\centering
\includegraphics[width=1.0\columnwidth, clip]{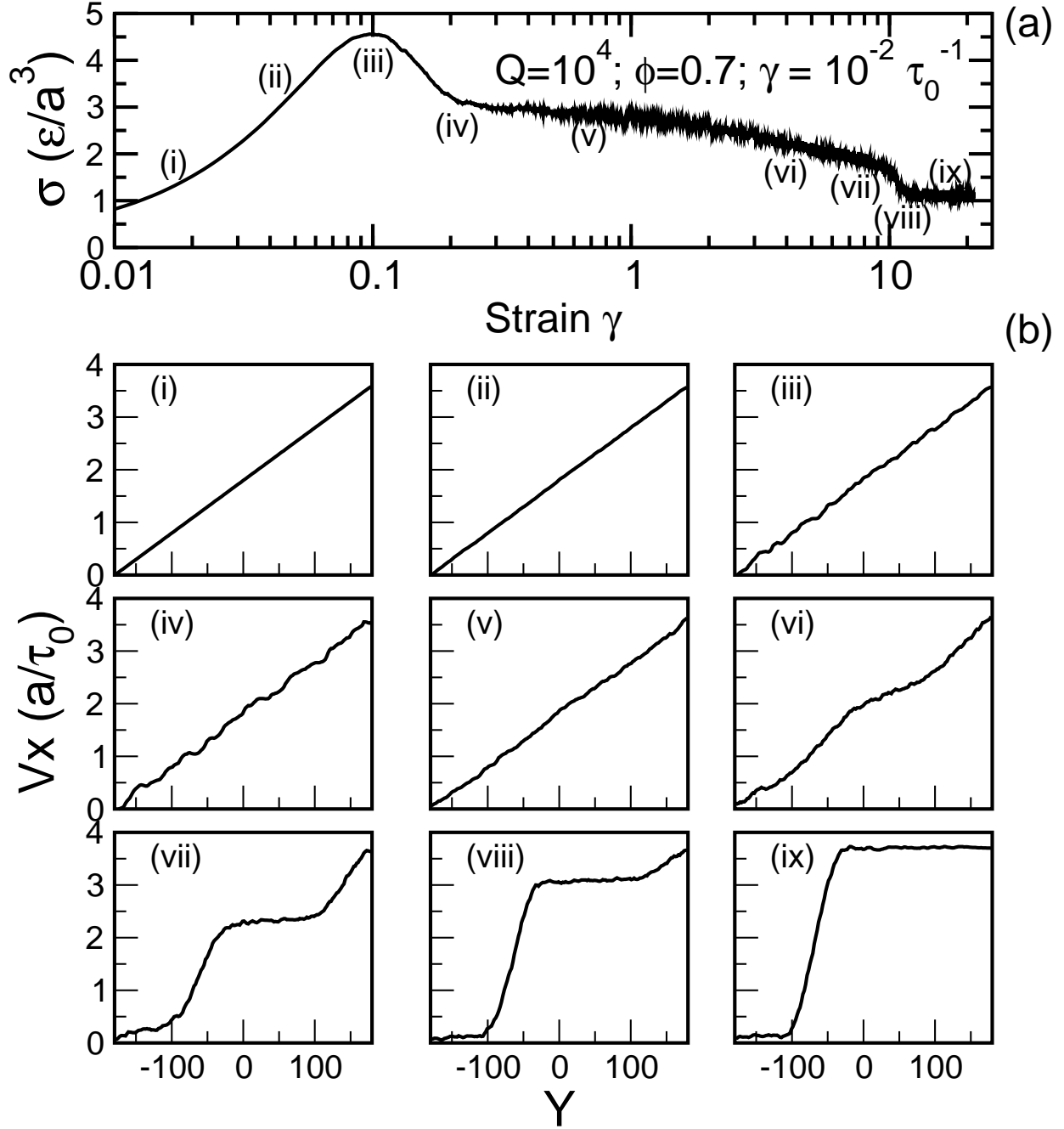}
\caption{(a) Load curve for a sample of size $L_x=L_z=22 a$ and $L_y=360 a$, sheared at a value of $Q=10^4$ (underdamped regime) and a shear rate $\dot{\gamma}=10^{-2}\tau_\mathrm{vib}^{-1}$. (b) Velocity profiles computed at different strains of the load curve, averaged over a strain window of $\Delta \gamma=0.02$.}
\label{SIfig1}
\end{figure}

\noindent{\bf Steady state: } In order to make sure that the sample, sheared at a chosen shear rate, has indeed reached a steady state we monitor the macroscopic load curve as well as the local flow profiles. In Fig.~\ref{SIfig1}(a) we show the load curve for $Q=10^4$, $L_x=L_z=22 a$, $L_y=360 a$, sheared at $\dot{\gamma}=10^{-2} \tau_\mathrm{vib}^{-1}$. In Fig.~\ref{SIfig1}(b) we show velocity profiles obtained at various strain values (shown on the load curve with roman numbers). We consider that the system has reached a steady state once the stress has relaxed to a stationary value, together with a stationary velocity profile, that shows complete coarsening into a single banded structure. We also monitor the local stress profile as shown in Fig.~\ref{fig1}(c) of the main manuscript, homogeneous across the system in steady state. The local stress is computed from the virial stress tensor averaged over the Voronoi volume defined around a particle. 

\noindent{\bf Herschel - Bulkley fitting parameters: } To obtain the Herschel-Bulkely (HB) fitting parameters for a finite temperature, over-damped system ($Q=1$), we have performed molecular simulations, over shear rates ranging between $\dot{\gamma}=10^{-5}\tau_\mathrm{vib}^{-1}$ to $\dot{\gamma}=10^{-1}$, of a system of $10K$ particles keeping all other parameters the same as the under-damped simulations. In order to extract over-damped HB parameters ($\sigma_Y$, $\kappa$ and $n$) at finite temperature we make sure that at each temperature (ranging between $T=0 \epsilon/k_B$ to $T=7.5 \epsilon/k_B$), the system has reached a steady state (monitoring the stress and the pressure) at each shear rate by shearing the system for large strain values ($\gamma$ upto $10$).\\

\noindent{\bf Minimum length to accommodate an instability in molecular simulations: }\\
In order to obtain the minimum length scale to accommodate the flow instability, using the eq.~(\ref{eq:cri-len}), one needs to compute the thermal conductivity $\lambda_T$, heat capacity $c_V$, the time associated with the relaxation of kinetic energy $\tau$, along with the derivative of stress with respect to temperature $\partial_T \sigma$ and to shear rate $\partial_{\dot{\gamma}}\sigma$. 

\begin{figure}[h]
\centering
\includegraphics[width=1.0\columnwidth, clip]{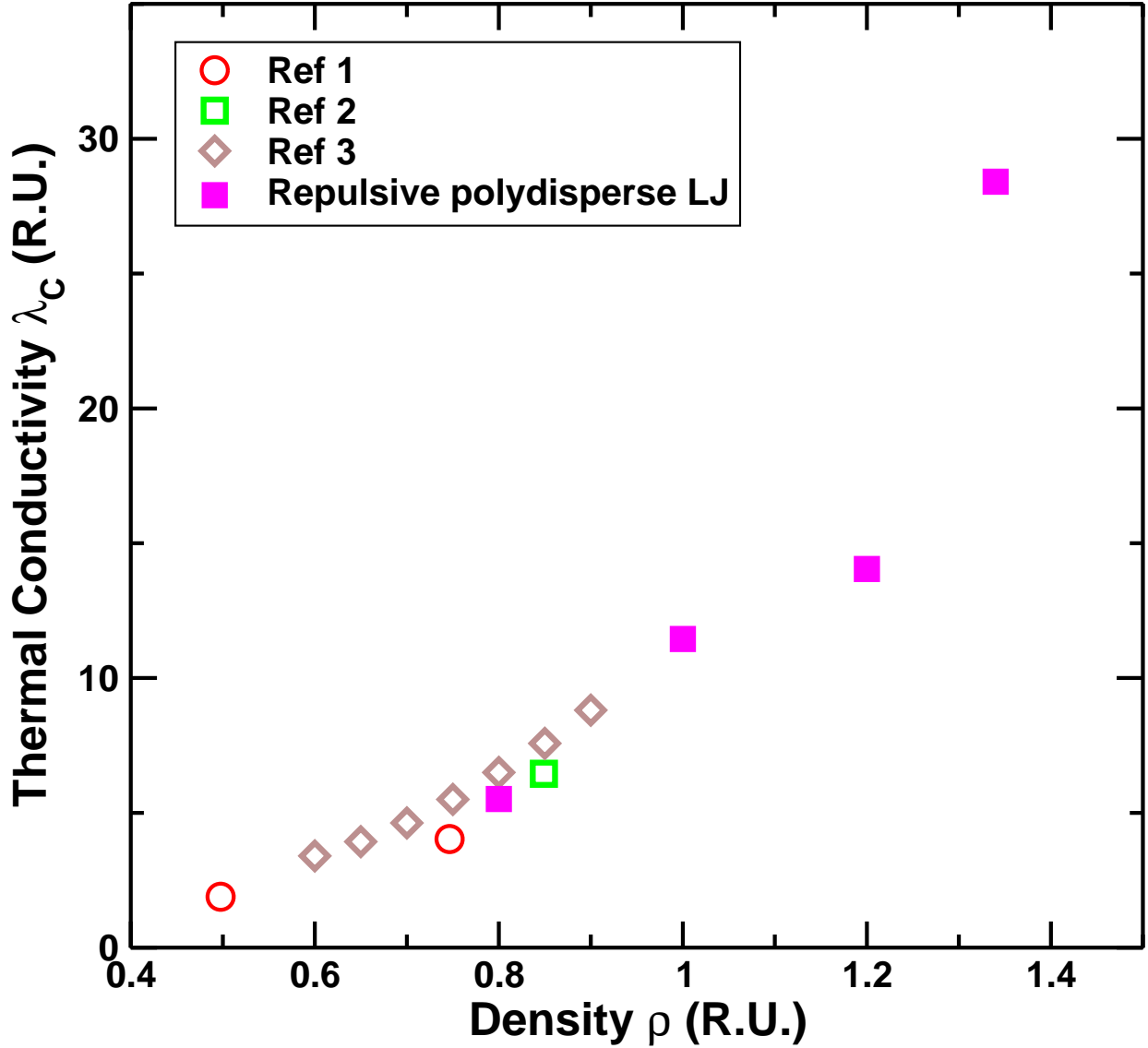}
\caption{Thermal conductivity $\lambda_T$ as a function of number density $\rho$. The opaque symbols represent our data computed for the system specified in manuscript. The open symbols corresponds to data available in literature, with Ref 1: Heyes and Powles, Molecular Physics (1998); Ref 2: Müller-Plathe, JCP (1997); Ref 3: Bugel and Galliero, Chemical Physics (2008).}
\label{SIfig2}
\end{figure}

\noindent {\it Thermal conductivity $\lambda_T$: } We compute $\lambda_T$ using the reverse non-equilibrium molecular dynamics method introduced by Muller-Plathe (JCP 106, 6082 (1997)). By imposing a heat flux $J$ on the system, we measure the resulting temperature gradient $\nabla T$ to compute the thermal conductivity from $J = -\lambda_T \nabla T$. We show in Fig.~\ref{SIfig2} the thermal conductivity (in LJ units) as a function of number density, along with a comparison of the data for $\lambda_T$ available in the literature. The system size used to calculate $\lambda_T$ is $N=97556$. For the purposes of computing $\ell_c$ in the manuscript, we have used $\lambda_T=30$, which corresponds to the volume fraction $\phi=0.7$.

\noindent {\it Heat capacity and relaxation time: } We have performed a slow quenching run to prepare the initial sample. From this data we compute the heat capacity as the slope of total energy vs.~temperature. To compute $\ell_c$, we use $c_V=3.0$ (LJ units). To compute the relaxation time associated with the kinetic energy, we monitor the dissipation of kinetic energy with time at a fixed dissipation constant corresponding to $Q=10^4$ and we find $\tau_\mathrm{KE} \equiv \tau \approx 350 \tau_\mathrm{vib}$.

\begin{figure}[h]
\centering
\includegraphics[width=0.8\columnwidth, clip]{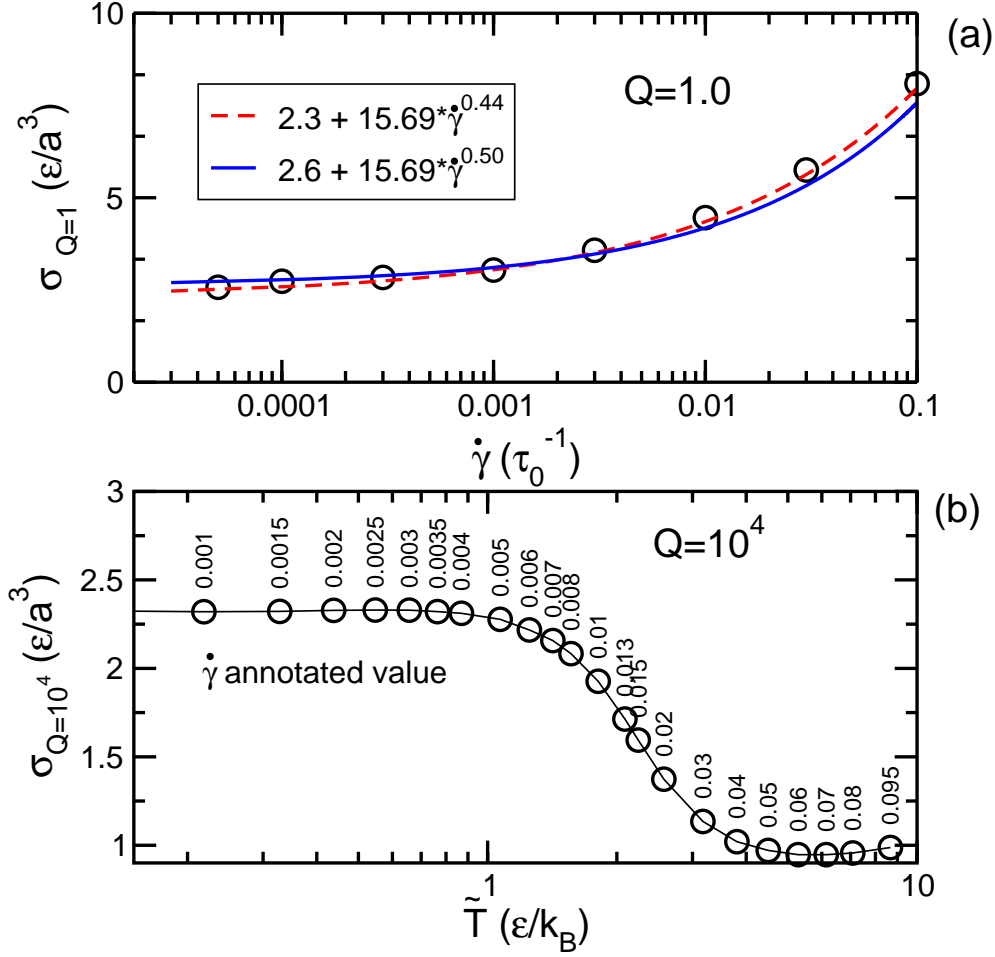}
\caption{(a) Flow curve data (open circle) obtained from overdamped simulations with HB fits (lines). (b) Measured stress against the measured kinetic temperature from underdamped simulations at different shear rates (annotated values).}
\label{SIfig3}
\end{figure}

\begin{figure}[h]
\centering
\includegraphics[width=0.8\columnwidth, clip]{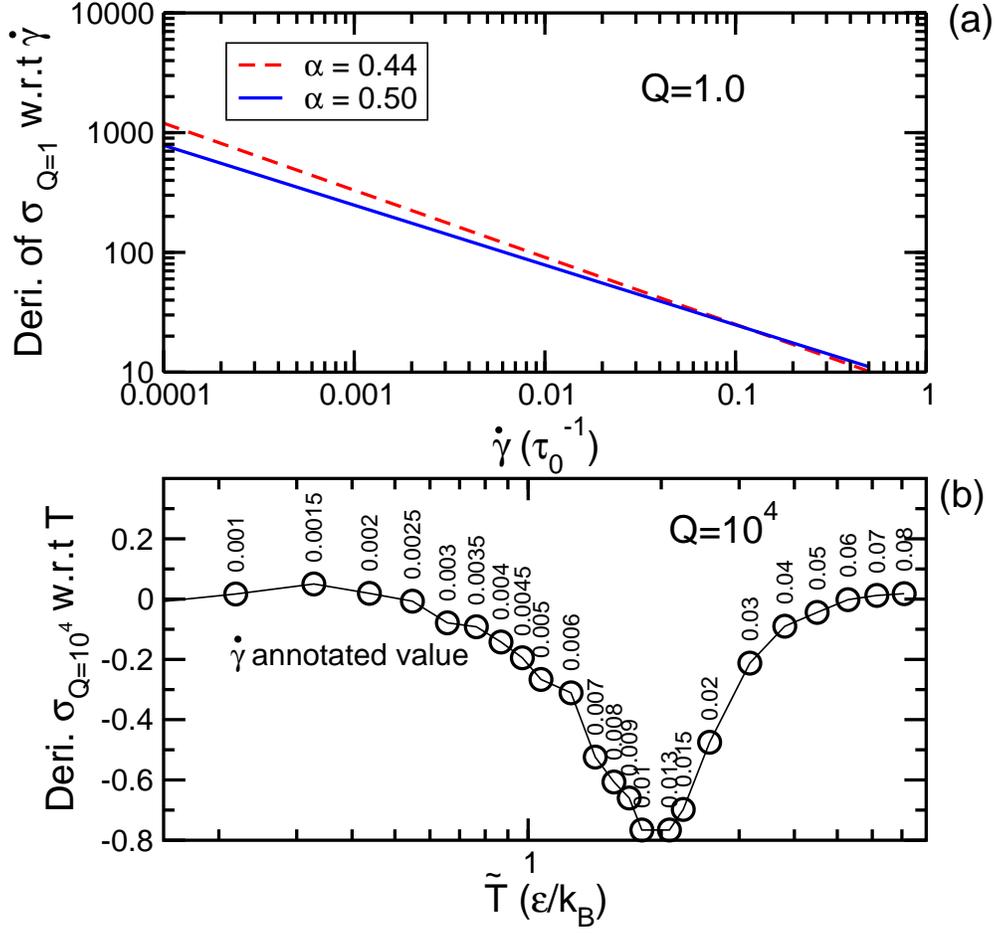}
\caption{(a) Derivative of HB equation with fit values obtained in Fig.~\ref{SIfig3}(a). (b) Numerical derivative of the data in Fig.~\ref{SIfig3}(b).}
\label{SIfig4}
\end{figure}

\begin{figure}[h!]
\centering
\includegraphics[width=0.8\columnwidth, clip]{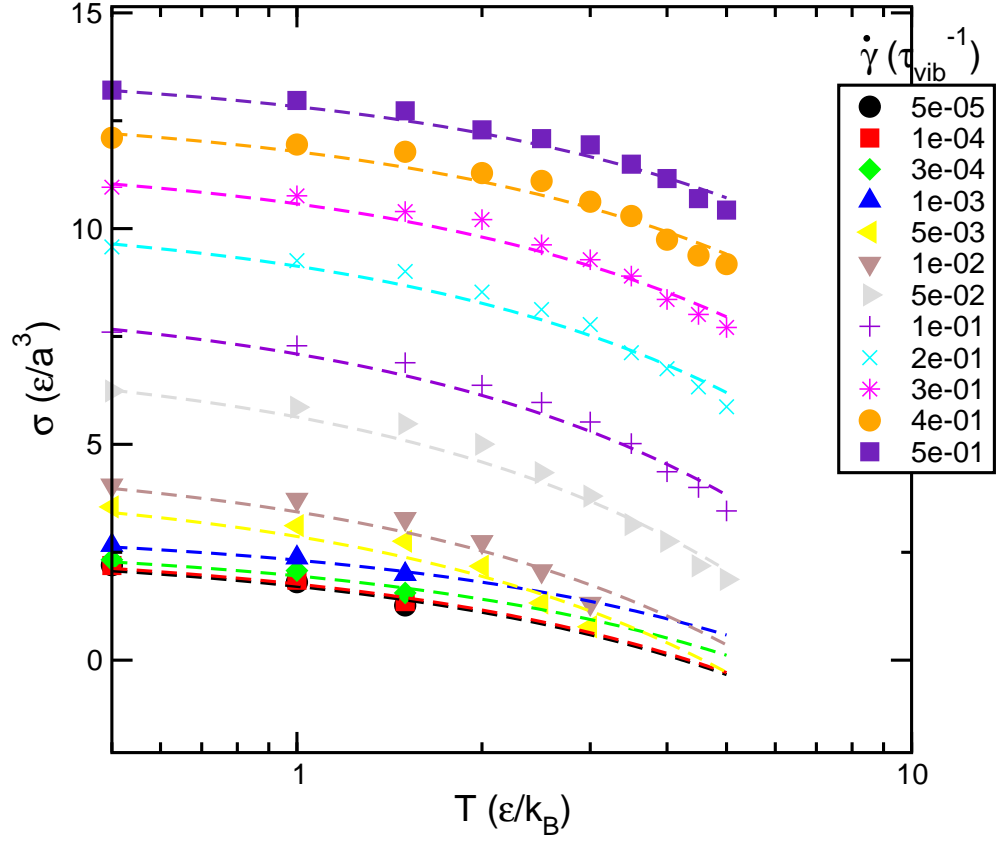}
\caption{Measured stress values (symbols), from overdamped, finite shear rate and temperature, against the applied temperature for different shear rates. The lines are the fits using the $T-\sigma$ relation from Ref.~\cite{chattoraj2010universal}.}
\label{SIfig5}
\end{figure}

\noindent{\bf Stress derivative: }The eq.~(\ref{eq:cri-len}) for the minimum length to accommodate the flow instability contains the partial derivative of stress w.r.t.~temperature ($\partial_{T}\sigma$) and w.r.t.~shear rate ($\partial_{\dot{\gamma}}\sigma$). One can obtain $\partial_{T}\sigma$ in two different ways. One way is to measure, at a given shear rate in a underdamped simulation ($Q=10^4$), the associated stress and the kinetic temperature and obtain a numerical derivative. The second way is to measure, at a given shear rate, the stress in the finite temperature simulations in overdamped simulation conditions ($Q=1$). The $\partial_{\dot{\gamma}}\sigma$ is obtained from the HB equation with fitting parameters obtained from the flow curve for $Q=1$. In Fig.~\ref{SIfig3}(a) we show the flow curve corresponding to $Q=1$ with the HB fit and in Fig.~\ref{SIfig3}(b) we show $\sigma$ {\it vs.} $\tilde{T}$ for $Q=10^4$ (with annotated values of the shear rates). In Fig.~\ref{SIfig4}(a) we show $\partial_{\dot{\gamma}}\sigma$ (for two different HB exponents) and in Fig.~\ref{SIfig4}(b) we show the numerical derivative $\partial_{T}\sigma$. In the small shear rate regime, the numerical derivative is very sensitive to minute changes in the simulation data. To overcome such difficulties, we use the $\sigma - T$ relationship derived in Ref. \cite{chattoraj2010universal} given by
\begin{equation*}
\sigma(\dot{\gamma},T)=\sigma(\dot{\gamma},T_0) - C_1 T^{2/3} \left [ ln\left ( C_2 T^{5/6}/\dot{\gamma} \right ) \right ]^{2/3}
\end{equation*}
\noindent where $C_1$ and $C_2$ were found to be shear rate independent in the Ref. 
\cite{chattoraj2010universal}. 
\begin{figure}[h!]
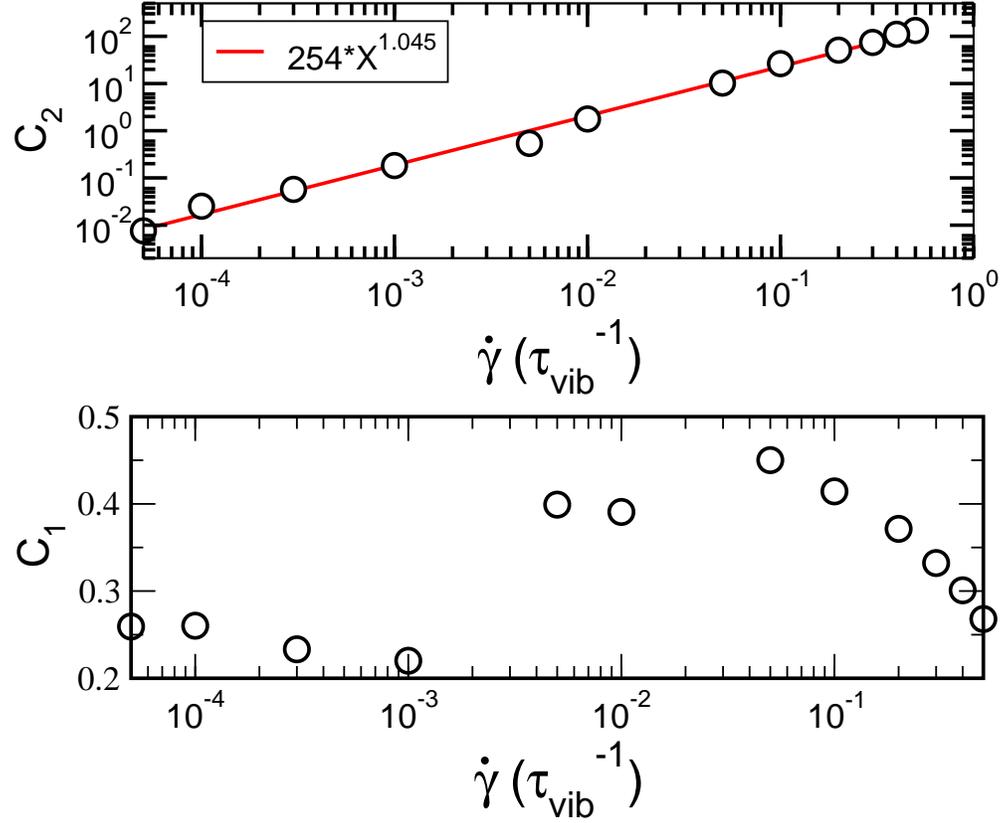

\centering
\includegraphics[width=0.8\columnwidth, clip]{SIFig6.eps}
\includegraphics[width=0.8\columnwidth, clip]{SIFig7.eps}
\caption{Upper panel: Fitting parameter $C_2$, obtained from Fig.~\ref{SIfig5} against the shear rate.
Lower panel: Fitting parameter $C_1$, obtained from Fig.~\ref{SIfig5} using $C_1/\dot{\gamma}={\mathrm const.}$, against the shear rate.}
\label{SIfig6}
\end{figure}
To obtain the $C_1$ and $C_2$, we perform shear simulations (at $Q=1$ and with N=10K particles) at finite temperature $T$ and finite shear rate and measure the stress in the system. In Fig.~\ref{SIfig5} we show $\sigma$ {\it vs.} $T$ for different shear rates. We fit the curves from the above relation and find that $C_1$ and $C_2$ are independent of shear rate only in a limited range. The coefficient $C_2/\dot{\gamma}$ is found to be constant (upper panel of Fig.~\ref{SIfig6}). Using this fact, we refit the curves in Fig.~\ref{SIfig5} and obtain $C_1$, shown in the lower panel of Fig.~\ref{SIfig6}. Using directly the coefficients obtained from the fits, we compute the derivative $\partial_{T}\sigma$. The error bars in Fig.~\ref{fig3} is obtained from partial derivatives computed numerically as well analytically with HB exponent $0.5$ and $0.44$.

\begin{figure}[h!]
\centering
\includegraphics[width=0.9\columnwidth, clip]{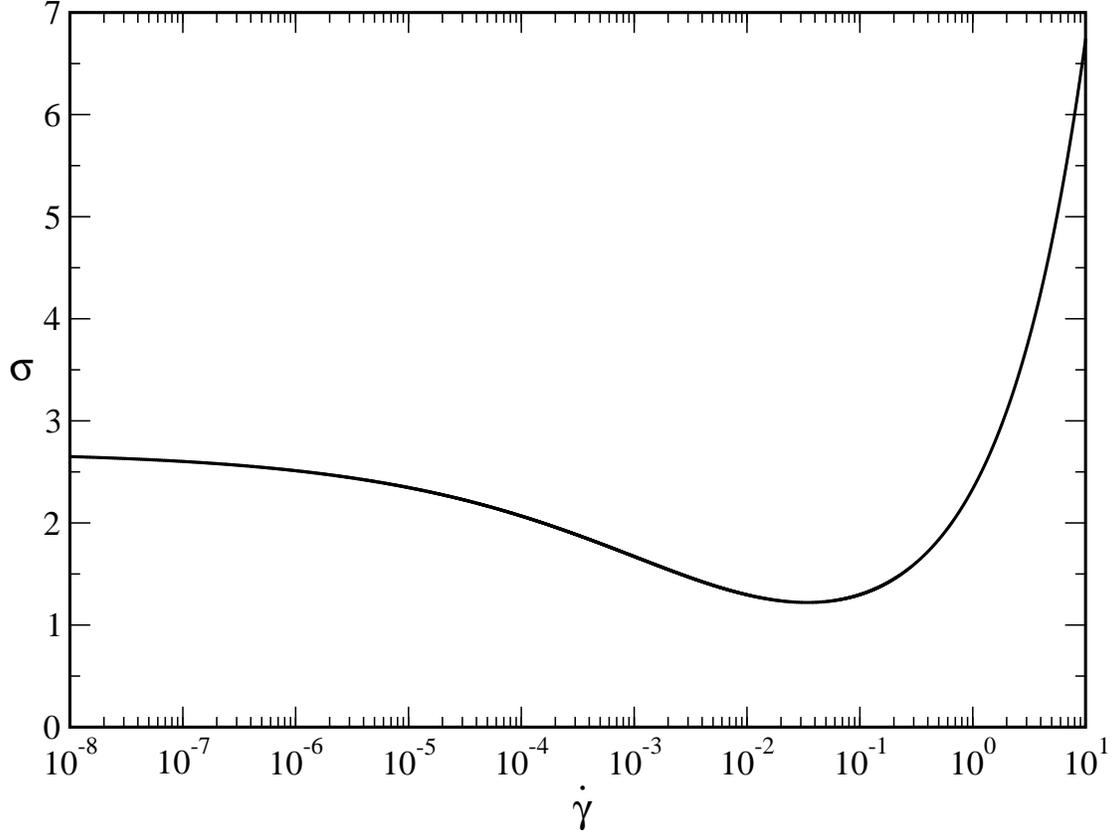}
\caption{Assumed constitutive relation for the analytical model obtained from solving equations (\ref{eq:T_steady_state}) and (\ref{eq:flowcurve}), using the implicit expression: $\sigma = \sigma_y + A\dot{\gamma}^n - B(\frac{\tau}{c_V} \sigma\dot{\gamma})^\alpha$ with the following parameters (in LJ units):
$c_V = 3$, 
$\rho = 1.337$, 
$\tau = 350$, 
$\sigma_y = 2.7$, 
$A = 12$, 
$B = 2.3$, 
$n = 0.5$, and
$\alpha = 0.3$
Since there is no general form of the temperature dependence of the flow-curve we implement the most simple form that reproduces the minimum in the flow-curve, $\dot{\gamma}_\mathrm{min}\simeq 5. 10^{-2}$ and $\sigma_\mathrm{min}\simeq 1.21$.}
\label{SIfigx}
\end{figure}

\noindent {\bf Flow curve for the continuum description :}
The assumed local constitutive relation used in the continuum description with the chosen model parameters is displayed in Fig.~\ref{SIfigx}.

\end{document}